# Negative and Positive Magnetoresistance in Bilayer Graphene: Effects of Weak Localization and Charge Inhomogeneity

Yung-Fu Chen, Myung-Ho Bae, Cesar Chialvo, Travis Dirks, Alexey Bezryadin, Nadya Mason

Department of Physics and Materials Research Laboratory, University of Illinois at Urbana-Champaign, Urbana, IL 61801-2902, USA

We report measurements of magnetoresistance in bilayer graphene as a function of gate voltage (carrier density) and temperature. We examine multiple contributions to the magnetoresistance, including those of weak localization (WL), universal conductance fluctuations (UCF), and inhomogeneous charge transport. A clear WL signal is evident at all measured gate voltages (in the hole doped regime) and temperature ranges (from 0.25 K to 4.3 K), and the phase coherence length extracted from WL data does not saturate at low temperatures. The WL data is fit to demonstrate that electron-electron Nyquist scattering is the major source of phase decoherence. A decrease in UCF amplitude with increasing gate voltage and temperature is shown to be consistent with a corresponding decrease in the phase coherence length. In addition, a weak positive magnetoresistance at higher magnetic fields is observed, and attributed to inhomogeneous charge transport.

## 1. Introduction

The realization of few layer [1-3] and then single layer graphite, so-called graphene [4,5], has paved the way for new science and technology. In particular, bilayer graphene has recently generated much excitement, as it can be considered a gapless semiconductor that is both chiral and non-relativistic. It consequently demonstrates unusual physics such as an anomalous quantum Hall effect [6,7] and strongly tunable tunneling of chiral carriers [8]. The interlayer coupling in bilayer graphene breaks the degeneracy of the π-band and anti-π-band, resulting in gapless quadratic dispersion at the K and K' points in the reciprocal lattice. The carriers have chirality and exhibit an overall Berry phase of $2\pi$ [9], in contrast to $\pi$ for single layer graphene [10]. Thus, unlike in single layer graphene [11], suppression of elastic backscattering is not expected. As a result, bilayer graphene can manifest enhanced elastic backscattering due to constructive interference between two time-reversed electron paths, or weak localization (WL) [12,13]. Weak localization occurs in disordered, phase coherent systems, and is evident as a zero-field peak in magnetoresistance. Studies of WL in bilayer graphene can thus elucidate quantum interference in the system. Previous work on WL in bilayer graphene has shown that the phase coherence length, $L_\varphi$, decreases with increasing temperature, likely due to the effect of electron-electron interactions [12]. However, because of saturation at low temperatures, the gate-voltage and temperature trends of $L_\varphi$ were difficult to determine.

In this paper, we describe magnetoresistance measurements on bilayer graphene over a relatively large span of magnetic field, allowing us to create a full picture of the magnetic field behavior as a function of temperature and gate voltage (carrier density). We observe and analyze signals of WL, universal conductance fluctuations (UCF), and a positive magnetoresistance background. We do not observe a saturation of the phase coherence length at low temperatures (down to 0.25 K). By comparing the temperature and gate voltage dependence of $L_\varphi$ with fits to theory, we demonstrate that the major phase decoherence source is inelastic electron-electron interactions. We also observe and explain a decrease in UCF amplitude with increasing temperature or decreasing carrier density. Finally, we observe positive magnetoresistance near the charge neutral point, which is explained by charge inhomogeneity in the system.

## 2. Sample Preparation and Measurements

The bilayer graphene samples were mechanically exfoliated onto highly doped Si substrates topped with 300 nm $SiO_2$. The thickness was determined by the color variation in optical microscopy, which had been calibrated by Raman and atomic force microscopy measurements. The electrodes were patterned by conventional electron beam lithography and electron beam evaporation of 3 nm Cr and 55 nm Au. The device shown in the inset of Fig. 1(a) consists of six electrodes on a piece of bilayer graphene, which allows measurements of longitudinal resistance $R_{xx}$ to determine 2D resistivity $\rho$, and Hall resistance $R_{xy}$ to determine carrier density $n_s$. The distance $L$ between two longitudinal electrodes and the minimum width $W$ of the sample are 6.4 and 8.0 μm, respectively. Although the sample shape is far from an ideal Hall bar geometry, the Hall data still seem valid, as the overall longitudinal magneto-voltage has no slope, indicating no mixing from $\rho_{xy}$; similarly, the slope of the Hall voltage does not seem to be affected by $\rho_{xx}$ (which only adds an offset, and dip due to WL, and some contributions from universal conductance fluctuations). The samples are measured in the hole-doping regime, as the contacts induce an electron-hole asymmetry [14]. Measurements were performed in a Helium-3 cryostat using standard ac lock-in techniques: an ac current of 3 – 40 nA was applied through two end electrodes and the resulting longitudinal voltage or Hall voltage was recorded as a function of gate voltage $V_g$ or magnetic field $B$ at various temperatures.

## 3. Results and Discussion

We determine the charge neutral point (zero average carrier density point) using $R_{xx}$ and Hall measurements as a function of $V_g$. A linear fitting of the Hall resistance as a function of magnetic field gives the carrier density $n_s$ (which is inversely proportional to the slope of the line; see Fig. 1(b)) [15]. By extrapolating the $n_s$ vs $V_g$ plot to zero carrier density (see inset of Fig 1(b)), we estimate the gate voltage at the charge neutral point, $V_{g,cnp} \sim 73$ V. This value compares well to the turn-over point in a plot of $R_{xx}$ as a function of $V_g$, as can be seen in Fig. 1(a) (we unfortunately could not attain much higher gate voltages because of the possibility of leakage currents). Also, as shown in Fig. 1(a), the gate voltage dependences of $R_{xx}$ and the residual $R_{xy}$ at $B = 0$ match very well, indicating the only effect related to the non-ideal Hall bar geometry in the sample is a field independent offset. The longitudinal resistance is also used to determine the mobility $\mu$ through equation $\mu = 1/n_s\rho = L/n_s R_{xx} W$, where $\rho$ is the sheet resistance. At $V_g = 0$ (i.e., far away from the charge neutral point), $n_s \sim 1.3 \times 10^{13}$ cm$^{-2}$, $\mu \sim 225$ cm$^2$/Vs, and the diffusion constant $D \sim 114$ cm$^2$/s.

Figure 2 shows $R_{xx}$ as a function of magnetic field at different temperature $T$ and $V_g$. For all curves the resistance drops quickly after a small magnetic field is applied, which results in a sharp peak near zero magnetic field. This behavior is consistent with WL, showing that the carriers maintain their phase over many mean free paths. At all $V_g$, as the temperature increases, the height of WL peak decreases and the width increases. The width of the WL peak is also carrier density dependent: as the hole density decreases the width of the WL peak increases (i.e., peaks in Fig. 2(d), near the charge neutral point, are wide compared to those in Fig. 2(a), which is away from the charge neutral point). These observations imply that the phase coherence length $L_\varphi$ of the sample is reduced by raising temperature and/or approaching the gate voltage corresponding to the charge neutral point.

WL theory in bilayer graphene [13] predicts that the magnetoresistance will follow:

$$\Delta\rho_{WL}(B) = -\frac{e^2\rho^2}{\pi h}[F(\frac{B}{B_\varphi}) - F(\frac{B}{B_\varphi + 2B_{iv}}) + 2F(\frac{B}{B_\varphi + B_{iv} + B_*})], \quad (1)$$

where $F(z) = \ln z + \psi(\frac{1}{2} + \frac{1}{z})$, $\psi$ is the digamma function, $B_{\varphi, iv, *} = \frac{\hbar}{4eD}L^{-2}_{\varphi, iv, *}$. $L_\varphi$ and $L_{iv}$ are the phase coherence length and elastic inter-valley scattering length, respectively. The term $L_*$ is related to the elastic intra-valley scattering length, $L_z$, and trigonal warping length, $L_w$, through the relation $L_*^{-2} = L_w^{-2} + 2L_z^{-2}$. Trigonal warping is related to the non-linearity of the dispersion away from the charge neutral point. From the scattering lengths we can extract the scattering rates $\tau^{-1}_{\varphi, iv, *}$ using the relation $L^2_{\varphi, iv, *} = D\tau_{\varphi, iv, *}$. It has been shown that the third term in Eq. (1) can be taken as negligible because the strong trigonal warping effect blows up the denominator [12,13]. In this case, inter-valley scattering is then required to create a WL effect, as it prevents the first two terms of Eq. (1) from cancelling.

Although Eq. (1) may be used to fit the magnetoresistance data for a very small range of magnetic fields, it does not fit at larger fields. In particular, in Fig. 2 it is clear that at larger fields away from the WL peaks the resistance increases with increasing magnetic field. This positive magnetoresistance becomes more pronounced near the charge neutral point (cf. Fig. 2(a) and 2(c)). The theory for bilayer graphene in Eq. (1)

does not predict positive magnetoresistance [13]: in single layer graphene a Berry phase of $\pi$ creates weak anti-localization, while the effect disappears in bilayers, which have a Berry phase of $2\pi$. The possible mechanism for this positive magnetoresistance in the bilayer sample is charge inhomogeneity in the system [16]. The conductivity near the charge neutral point may be dominated by charge disorder, which creates electron and hole puddles in the sample [17]. Such a two-fluid system having equal mobility for both types of carriers—and thus a net drift velocity perpendicular to the applied current at finite magnetic field—can yield positive magnetoresistance [16,18]. The two-fluid situation is expected to be more pronounced as the average carrier density approaches zero; therefore the positive magnetoresistance should be more noticeable near the charge neutral point. The resistivity correction due to charge inhomogeneity in the system is proposed by [16] to be:

$$\Delta\rho_{CI}(B) = (\sigma_{xx,1} + \frac{\sigma_{xx,0}}{(1+(\mu B)^2)^{\frac{1}{2}}})^{-1} - (\sigma_{xx,1} + \sigma_{xx,0}), \qquad (2)$$

where $\sigma_{xx,0} + \sigma_{xx,1}$ is the conductivity at $B = 0$. $\sigma_{xx,1}$ is a phenomenological fitting parameter and is necessary to obtain reasonable fits to the data [16]. Although positive magnetoresistance due to charge inhomogeniety has been demonstrated in single-layer graphene [16], it has not been seen previously in bilayer devices. The effect should be weaker in bilayer devices, as the electron-hole puddles are less robust. This may explain why the positive magnetoresistance we observe is weaker than that seen in Ref. [16]. However, the range of charge densities over which the positive magnetoresistance is seen, as well as the strengthening of the effect near the charge neutral point, are consistent with previous results and predictions for charge inhomogeneity effects. The asymmetry of magnetoresistance observed at $T \sim 260$ mK depends on the direction of field sweep, but not rate, and the origin is still unknown.

The bilayer graphene magnetoresistance data can now be fitted by combining Eq. (1) and Eq. (2) to obtain $\rho_{xx}(B) = \rho_{xx,B=0} + \Delta\rho_{WL} + \Delta\rho_{CI}$, where $R_{xx} = (L/W)\rho_{xx}$. The fits are plotted in Fig. 2 as thin solid lines. The theory fits the data well with $L_\varphi$, $L_{iv}$, $\mu$ and $\sigma_{xx,1}$ as four free parameters. As an example of fitting values, for $V_g = 40$ V and $T = 269$ mK (the black curve in Fig. 2(b)), we find $L_\varphi = 0.30$ μm, $L_{iv} = 0.16$ μm, $\mu = 2700$ cm$^2$/Vs, $\sigma_{xx,1} = 5.6$ μS [19] From the fittings we always find that $L_\varphi > L_{iv}$ at low temperatures, and that inter-valley scattering has thus acted to restore the WL peak [20]. Although $L_{iv}$ does not exhibit clear temperature or gate voltage dependence, consistent with what has been seen previously in bilayer graphene [12], $L_\varphi$ has systematic dependence on both temperature and gate voltage. For a two-dimensional conductor with electron phase coherence limited by electron-electron interaction through inelastic Nyquist scattering [21], $L_\varphi$ is described by [22]:

$$L_\varphi^{-2} = \beta \frac{k_B T}{D} \frac{\ln g}{g} \tag{3}$$

where $g = \sigma/(e^2/h)$ is the normalized conductivity, $k_B$ is Boltzmann constant, and $\beta$ is a numerical constant on the order $\sim 1$. Nyquist scattering is the scattering of electrons off the fluctuating electromagnetic fields generated by all other electrons. In Fig. 3(a) we plot $L_\varphi$ as a function of $1/\sqrt{T}$ at different $V_g$. Linear fits are shown as thin lines, demonstrating that $L_\varphi$ fits very well to $1/\sqrt{T}$ for all $V_g$, as expected for Nyquist scattering. Eq. (3) also predicts that $L_\varphi$ increases with decreasing $V_g$ (increasing hole carrier density), since the normalized conductivity $g$ of the sample increases with carrier density. This trend is evident in Fig. 3(a). The decrease of $L_\varphi$ with carrier density has been observed before in single layer graphene [20,23,24] and may be influenced by the formation of puddles of different types of carriers [17,20].

The decrease of the fitting slope in Fig. 3(a) with increasing $V_g$ (decreasing hole density $n_s$) is also explained by Eq. (3): the smaller the normalized conductivity $g$, the smaller the slope. In Fig. 3(b) we plot the slopes of the line fits in Fig. 3(a) (solid squares), and compare them with the slopes calculated from Eq. (3) using $\beta$ = 1.5 and $g$ and $D$ extracted from Fig. 1(a) (dotted line). The two curves agree well in terms of trend and magnitude. This is further confirmation that the major phase source of decoherence in the bilayer graphene sample is electron-electron interaction through inelastic Nyquist scattering, as suggested by Ref. [12]. The main difference between the results in our bilayer sample and those in Ref [12]—and the reason we can extract more data to observe trends—is that we do not observe the saturation of $L_\varphi$ at low temperatures (at least down to 250 mK). This saturation has been explained by the fact that $L_\varphi$ approaches the sample size. In our case, the sample size is on the order of 10 μm in both length and width, while $L_\varphi$ < 1.5 μm. The expected $L_\varphi$ saturation due to sample size may occur at much lower temperatures.

Fluctuations in the longitudinal magnetoresistance are also evident in Fig. 2 and can be attributed to universal conductance fluctuations (UCF). UCF are due to interference between different electron (or hole) trajectories in a mesoscopic phase-coherent sample that are altered randomly by a magnetic field. The fluctuations are aperiodic and have a magnitude on the order of $e^2/h$. In Fig. 2 it can be seen that as $T$ or $V_g$ increase, the fluctuation amplitude and frequency decrease [25]. The root mean square of the fluctuation amplitude, $G_{fluc,rms} = (G_{fluc,var})^{1/2}$, $G_{fluc,var} = \langle(G_{fluc} - \langle G_{fluc}\rangle)^2\rangle$, is plotted in Fig. 3(c). It is evident that UCF at low temperatures or away from the charge neutral point have amplitudes close to (though smaller than) $e^2/h$; however, these fluctuations are suppressed as $T$ increases or $V_g$ moves toward the charge neutral point. This phenomenon has been recently shown in bilayer and tri-layer graphene [26], as well as in single layer devices [24,27]. In Ref [24] we discussed how the amplitude suppression is due to $L_\varphi$ decreasing and is expected in a device where $L_\varphi$ is smaller than sample size. A similar reasoning can explain why $G_{fluc,rms}$ decreases in the bilayer sample when $L_\varphi$ decreases with $T$ or $V_g$. The fact that the UCF amplitudes are generally smaller than $e^2/h$ can also be explained by a sample size greater than $L_\varphi$. This is consistent with the lack of phase coherence length saturation at low temperatures, discussed above in the context of the WL analysis.

### 4. Summary

In summary, we have examined magnetoresistance in bilayer graphene as functions of temperature and carrier density. The magnetoresistance curves fit well to the predicted theory of WL in bilayer graphene in combination with a theory of charge inhomogeneity. The former accounts for sharp negative magnetoresistance peaks, while the latter explains a broader positive magnetoresistance. From the WL data and fitting, we demonstrate that the major phase decoherence source is electron-electron interactions via inelastic Nyquist scattering. We do not observe a saturation of the phase coherence length at low temperatures, probably because the sample size is greater than the phase coherence length in this regime. Finally, we observe a decrease in the amplitude of universal conductance fluctuations with increasing temperature and gate voltage, due to a corresponding loss of phase coherence.

Figure Captions:

Figure 1. Bilayer graphene device characterization. (a) Longitudinal resistance $R_{xx}$ at 4.3 K (solid curve) and the residual $R_{xy}$ at $B = 0$ (solid squares, data from (b)) as a function of $V_g$. Inset of (a) shows an optical image of the measured device. The contrast between graphene piece and substrate shows that the piece is bilayer graphene. (b) Hall magnetoresistance at different $V_g$ at 260 mK ($V_g$ varies from -40 V to 40 V in steps of 20 V from top to bottom). $dR_{xy}/dV_g$ is used to determine the carrier density at each $V_g$. Inset of (b) plots carrier density (in unit of $10^{12}$ cm$^{-2}$) as a function of $V_g$ (positive carrier density indicates that the carrier are holes). The linear fit in the inset shows that the $V_g$ induced carrier density is $1.79 \times 10^{11}$ cm$^{-2}$V$^{-1}$ [28].

Figure 2. Longitudinal magnetoresistance at different $V_g$ and $T$. The charge neutral point occurs at $V_g \sim 73$ V. (a) $V_g = -40$ V. (b) $V_g = 40$ V. (c) $V_g = 50$ V. (d) $V_g = 70$ V. WL peaks near zero magnetic field are observed at all examined $V_g$ and $T$. The data is fit to a combined theory of bilayer graphene WL theory proposed by Kechedzhi *et al*. [13] (without the third term in Eq. (1)) and an equal mobility theory proposed by Cho *et al*. [16]. Inset of (a): smaller magnetic field range (-0.1 ~ 0.1 T) of longitudinal magnetoresistance at $V_g = -40$ V and $T = 0.256$ K.

Figure 3. (a) $L_\varphi$ as a function of $1/\sqrt{T}$ at different $V_g$, obtained from WL fitting of magnetoresistance. The lines are linear fitting for each $V_g$, showing that $L_\varphi$ follows $1/\sqrt{T}$ dependence well. (b) The slope of the linear fitting in (a) as a function of $V_g$. The dotted line is the calculated slope according to $L_\varphi = (D\tau_\varphi)^{1/2} = (\dfrac{D}{\beta k_B T \ln g / g})^{1/2}$ with $\beta =$

1.5. (c) Root-mean-square of $G_{\text{fluc}}$ (in unit of $e^2/h$) as a function of $T$ (with the same legend as (a)).

Figure 1

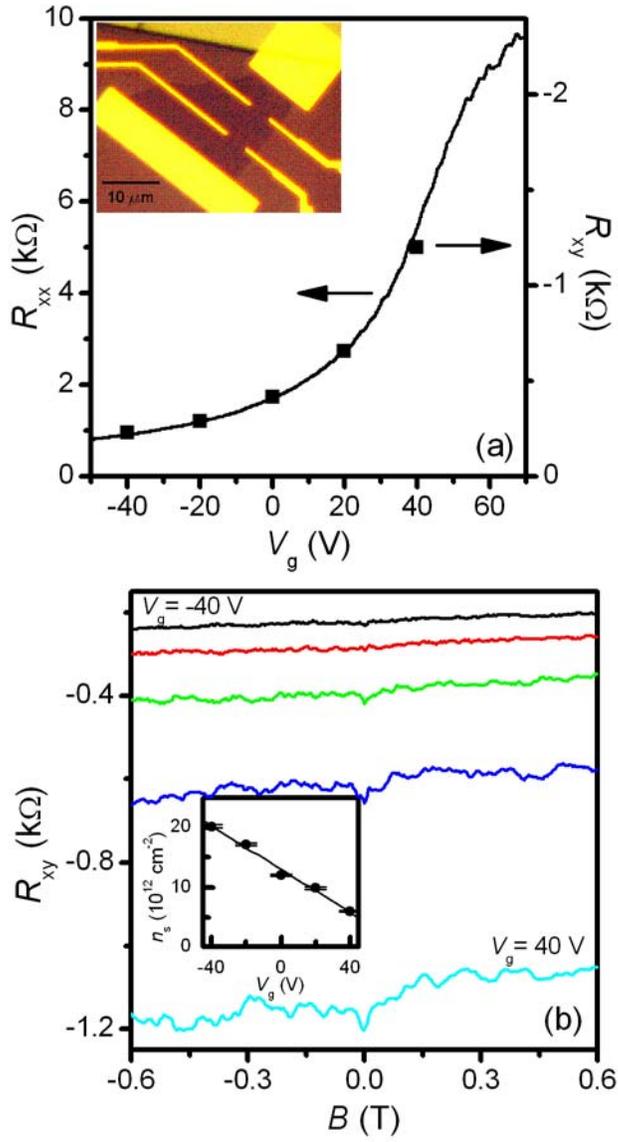

Figure 2

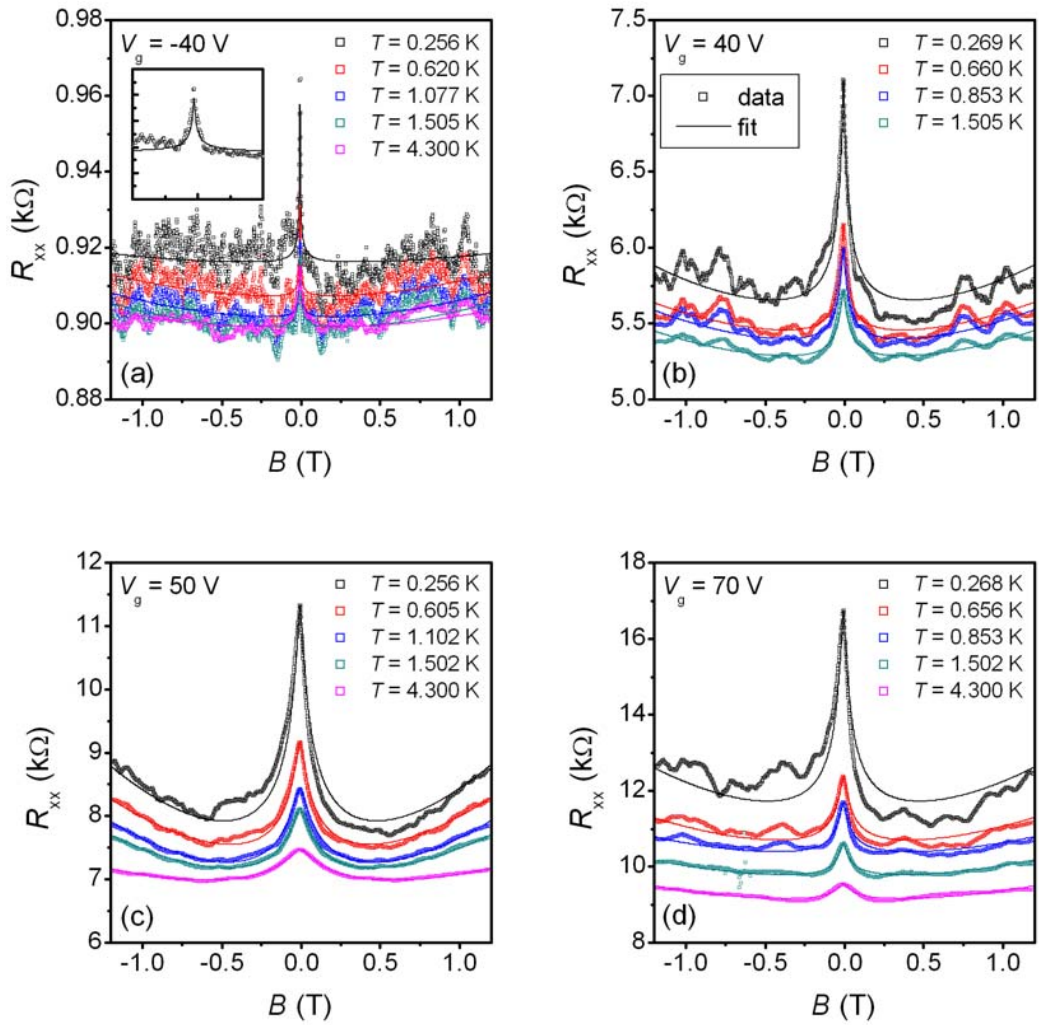

Figure 3

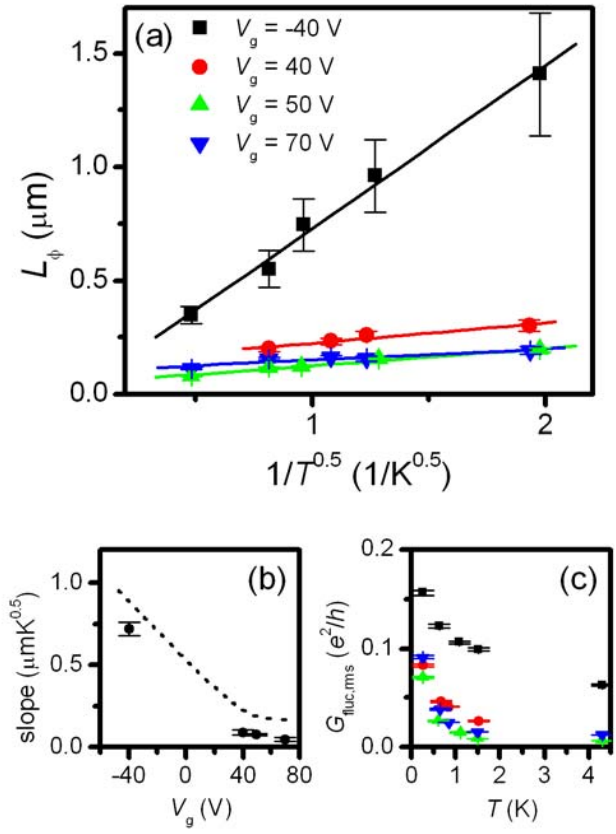